\documentclass[global,twocolumn,final]{svjour}  
\usepackage{graphicx}
\usepackage{apalike}

\journalname{Biological Cybernetics}

\begin{document}

\title{Edge vulnerability in neural and metabolic networks}
\author{Marcus Kaiser \thanks{{\it Correspondence to}: M. Kaiser (\email{m.kaiser@iu-bremen.de})} \and Claus C. Hilgetag}
\institute{International University Bremen, School of Engineering and Science, Campus Ring 6, 28759 Bremen, Germany}

\date{\today}

\maketitle

\begin{abstract}
Biological networks, such as cellular metabolic pathways or networks of corticocortical connections in the brain, are intricately organized, yet remarkably robust toward structural damage.  Whereas many studies have investigated specific aspects of robustness, such as molecular mechanisms of repair, this article focuses more generally on how local structural features in networks may give rise to their global stability. In many networks the failure of single connections may be more likely than the extinction of entire nodes, yet no analysis of edge importance (edge vulnerability) has been provided so far for biological networks. We tested several measures for identifying vulnerable edges and compared their prediction performance in biological and artificial networks. Among the tested measures, edge frequency in all shortest paths of a network yielded a particularly high correlation with vulnerability, and identified inter-cluster connections in biological but not in random and scale-free benchmark networks. We discuss different local and global network patterns and the edge vulnerability resulting from them.
\end{abstract}


\maketitle

\section{Introduction}
Extensive evidence shows that biological networks are remarkably robust against damage of their nodes as well as links among the nodes. For example, Parkinson disease only becomes apparent after a large proportion of pigmented cells in the substantia nigra are eliminated \cite{Damier1999}, and in spinal cord injuries in rats, as little as 5\% of the remaining cells allow functional recovery \cite{You2003}. Many metabolic networks, as well, were found to be robust against the knockout of single genes. This feature is both due to the existence of duplicate genes as well as alternative pathways which ensure that a certain metabolite can still be produced using the undamaged parts of the network \cite{Wagner2000}. 

Approaches of network analysis have been used to investigate various types of real-world networks, in which persons, proteins, brain areas or cities are considered nodes; and functional interactions or structural connections are represented as edges of the network \cite{Strogatz2001}. Many such systems display properties of small-world networks \cite{Watts1998}, with clustered local neighborhoods and short characteristic paths (or average shortest paths, ASP). Also, some networks possess more highly connected nodes, or hubs, than same-size random networks, leading to a power-law degree ({\it scale-free}) distribution of edges per node, where the probability for a node possessing $k$ edges follows $k^{-\gamma}$ \cite{Barabasi1999}. Such scale-free networks are error-tolerant towards random elimination of nodes, but react critically to the targeted elimination of highly connected nodes \cite{Barabasi2000a}. It has been shown that cerebral cortical networks in the cat and macaque monkey brain display a similar behavior \cite{Kaiser2004e}. While networks may possess features of both  small-world and scale-free organization, the two topologies are not necessarily identical. 

Whereas previous studies explored the impact of lesioning network nodes \cite{Barabasi1999}, the effect of {\it edge} elimination in biological networks has not yet been investigated. How can edges that are integral for the stability and function of a network be identified? In some systems, for instance, transportation or information networks, functional measures for the importance of edges, such as flow or capacity, are available. For many biological networks, however, such measures do not exist. In brain connection networks, for example, a projection between two regions may have been reported, but its structural and functional strength is frequently unknown or not reliably specified \cite{Felleman1991}, and its functional capacity may vary depending on the task \cite{Buchel1997}. Similarly, in biochemical networks, reaction kinetics are often highly variable, or generally unknown \cite{Schuster1994,Stelling2002}. However, the analysis of a network's structural organization may already provide useful information on the importance of individual nodes and connections, by identifying local features and investigating their importance for global network structure and function. Specifically, we are interested in these questions: Which structural patterns can be identified in a network? Have biological networks specific features, compared to artificial networks? As examples for biological networks we analyzed cortical fiber networks, metabolic networks as well as protein interaction networks and, for comparison, also an artificial network, the German highway system. 

In our analyses we tested the effect of eliminating single edges from networks. For some of the studied networks, such as cortical connectivity, structural damage might also result in multiple lesions of connections or regions. Such lesions can lead to network fragmentation, and may be theoretically analyzed within the framework of edge or vertex cut sets \cite{Sporns2002}. A computational algorithm to predict the effect of multiple network perturbations (or lesions) was presented as part of the MSA (Multi-perturbation Shapley value Analysis) approach \cite{Keinan2004}. In the present article, however, we focus on the impact of eliminating single edges, to infer network patterns and their vulnerability.

Two established parameters for characterizing networks are the average shortest path (ASP) and the clustering coefficient (CC). The ASP is the average number of edges that have to be crossed in order to reach one node from another. For a network with $N$ nodes, it is calculated as the average of all existing shortest paths:\\
\begin{equation}\label{asp}
ASP = \overline{d(i,j)}   \ \ \ \ \ \mbox{with $i\ne j$ and $d(i,j)\ne \infty$},
\end{equation}
where $d(i,j)$ is the number of edges of the shortest path between nodes $i$ and $j$. Infinite distances between unconnected nodes, which occur after network fragmentation, are excluded from the average calculation. We used the deviation of the ASP before and after edge elimination as a measure of network damage.

The clustering coefficient of a node $v$ with $k_v$ adjacent nodes (neighbors) is defined as the number of edges existing among the neighbors, divided by the number $k_v^2 - k_v$ of all possible edges among the neighbors \cite{Watts1999}. We use the term clustering coefficient as the average clustering coefficient for all nodes of a network. Networks with an ASP comparable to randomly connected networks, but with much higher clustering coefficient, are called {\it small-world} networks \cite{Watts1998}. These systems exhibit network clusters, that are regions, in which many interconnections exist within a cluster, but only few connections run between clusters. Various kinds of networks, such as electric power grids and social networks, display small-world properties \cite{Watts1998}. In addition, neural networks of {\it C. elegans} and cortical networks of the cat and macaque were also shown to be small-world networks \cite{Watts1998,Hilgetag2000b,Sporns2000} and to exhibit a clustered architecture \cite{Hilgetag2000b}.

\section{Materials and methods}

\subsection{Investigated biological and artificial networks}

\subsubsection{Brain networks}
We investigated long-range fiber projections in the cat and macaque brain. Nodes were brain regions or areas (e.g., V1), and edges were fiber connections between them. We considered connectivity data for a non-human primate, the Macaque monkey (73 nodes; 835 edges; density 16\%), and the cat (55 nodes; 891 edges; density 30\%) \cite{Scannell1995,Scannell1999,Young1993}. In both species, the data included connections between cortical regions, as well as a few subcortical structures (e.g., the amygdala) and regions of entorhinal cortex. These networks possess properties of small-world networks \cite{Hilgetag2000b}, and also show similar response to attack as scale-free networks \cite{Kaiser2004e}. The clustering coefficient of these networks ranged between 40 to 50\% (cf. Table \ref{corrtable}).

\subsubsection{Protein-protein interactions} 
As an example of biochemical networks, we examined the protein-protein interactions of the {\it S. cerevisiae} yeast proteome \cite{Jeong2001}. The data consisted of 1,846 proteins and 2,203 distinct functional relationships among them, forming 4,406 unidirectional edges in the network graph (data from http://\-www.nd.edu/\-$\sim$networks/\-database/). With a value of 6.8\%, the clustering coefficient was considerably smaller than for brain networks. As a further difference, the protein interactions network consisted of 149 disconnected components. The main component contained 79\% of the proteins, and the remaining components mostly were composed of only one pair of proteins. As shown before \cite{Jeong2001}, connections in this network are also distributed in a scale-free fashion. However, the yeast two-hybrid method yielding protein interaction data \cite{Ito2001,Gavin2002} produces many artifacts \cite{Kitano2002} which might have influenced prediction results.

\subsubsection{Metabolic networks} 
Cellular metabolic networks of different species were analyzed. Nodes here were metabolic substrates, and edges were considered as reactions \cite{Ravasz2002} (Data at http://www.nd.edu/\-$\sim$networks/\-database/). For bacteria, we investigated the metabolism of {\it E. coli} with 765 metabolites and 3,904 reactions. For eukaryotes, the data for {\it Arabidopsis thaliana} (299 metabolites and 1,276 reactions) and the yeast {\it S. cerevisiae} (551 metabolites and 2,789 reactions) were examined.

\subsubsection{Transportation network}
Data for the German highway (Autobahn) system were explored as a comparative example of man-made transportation networks. The network consisted of 1,168 location nodes (that is, highway exits) and 2,486 road links between them (Autobahn-Informations-System, AIS, from http://www.bast.de). Only highways were included in the analysis, discarding smaller and local roads ('Bundesstrassen' and 'Landstrassen'). Parking and resting locations were also excluded from the set of nodes. Furthermore, multiple highway exits for the same city were merged to one location representing the city as a single node of the network graph.

\subsubsection{Benchmark networks}
Twenty random networks with 73 nodes and comparable density as the Macaque network were generated. Moreover, twenty scale-free networks with 73 nodes and equivalent density were grown by preferential attachment \cite{Barabasi1999}, starting with an initial matrix of 10 nodes.

In addition to the random and scale-free networks, which consisted of only one cluster, networks with multiple clusters\label{multibench} were considered. Twenty networks were generated with 72 nodes in order to yield three clusters of the same size with 24 nodes each. Connections within the clusters were distributed randomly, and six inter-cluster connections were defined to mutually connect all clusters. Average density of these networks was again similar to the Macaque data. \\

\begin{table*}
\caption[smallcaption]{Density, clustering coefficient CC, average shortest path ASP and correlation coefficients $r$ for different vulnerability predictors of the analyzed networks (the index refers to the number of nodes). Tested prediction measures were the product of degrees (PD), absolute difference of degrees (DD), matching index (MI), and edge frequency (EF). }
\label{corrtable}
\begin{tabular}{llll|llll}
\hline\noalign{\smallskip}
 & Density & CC & ASP & $r_{PD}$ & $r_{DD}$ & $r_{MI}$ & $r_{EF}$ \\
\noalign{\smallskip}\hline\noalign{\smallskip}
Macaque$_{73}$ & 0.16 & 0.46 & 2.2 & 0.10$^{**}$ & 0.57$^{**}$ & -0.40$^{**}$ & 0.84$^{**}$ \\
Cat$_{55}$ & 0.30 & 0.55 & 1.8 & 0.08$^{*}$ & 0.48$^{**}$ & -0.34$^{**}$ & 0.77$^{**}$ \\
AT$_{299}$ (metabolic) & 0.014 & 0.16 & 3.5 & 0.04 & 0.09$^{**}$ & -0.11$^{**}$ & 0.74$^{**}$ \\
EC$_{765}$ (metabolic) & 0.0067 & 0.17 & 3.2 & 0.31$^{**}$ & 0.38$^{**}$ & -0.15$^{**}$ & 0.75$^{**}$ \\
SC$_{551}$ (metabolic) & 0.0092 & 0.18 & 3.3 & 0.11$^{**}$ & 0.22$^{**}$ & -0.04 & 0.74$^{**}$ \\ 
SC$_{1846}$ (protein interactions) & 0.0013 & 0.068 & 6.8 & 0.24$^{**}$ & 0.02 & -0.14$^{**}$ & 0.60$^{**}$ \\
German highway$_{1168}$ & 0.0018 & 0.0012 & 19.4 & 0.19$^{**}$ & 0.06$^{**}$ & -0.04 & 0.63$^{**}$ \\

Random$_{73}$ & 0.16 & 0.16 & 1.7 & 0.02 & 0.06 & 0.00 & 0.03 \\
Scale-free$_{73}$ & 0.16 & 0.29 & 2.0 & 0.03 & 0.08 & -0.01 & 0.03 \\
\noalign{\smallskip}\hline
\end{tabular} \\
{\small $^{*}$  Significant Pearson Correlation, 2-tailed 0.05 level.}\\
{\small $^{**}$ Significant Pearson Correlation, 2-tailed 0.01 level.}\\
\end{table*}

\begin{figure}
\includegraphics{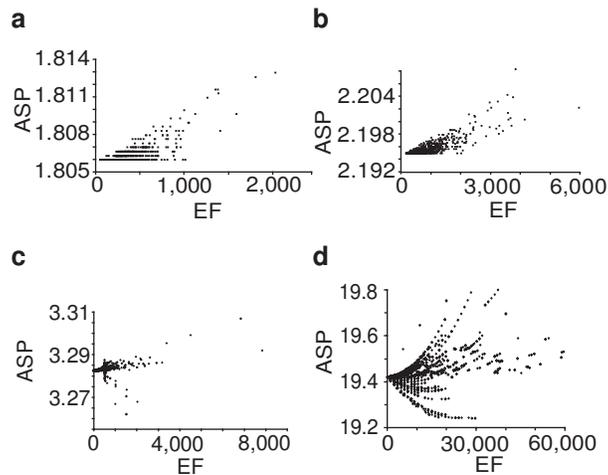} 
\caption{Frequency of edges in the all-pairs shortest paths and resulting network damage after elimination. {\bf a} Cat brain connectivity and primate (Macaque) brain connectivity {\bf b} show a strong correlation with damage. {\bf c} Metabolic network of {\it S. cerevisiae}. {\bf d} German highway system. Decreases of ASP were caused by eliminated cut-edges, leading to a separation of the network. \label{lincorr}}
\end{figure}

\subsection{Methods for detecting important connections}
We tested four candidate measures for predicting vulnerable edges in networks. All algorithms were programmed in Matlab (Release 12, MathWorks Inc., Natick) as well as implemented in C for larger networks. Links were analyzed as directed connections for all networks. 

First, the product of the degrees (PD) of adjacent nodes was calculated for each edge. A high PD indicates connections between two hubs which may represent potentially important network links. 

Second, the absolute difference in the adjacent node degrees (DD) of all edges was inspected. A large degree difference signifies connections between hubs and more sparsely connected network regions which may be important for linking central with peripheral regions of a network. 

Third, the matching index (MI) \cite{Hilgetag2002b} was calculated as the number of matching incoming and outgoing connections of the two nodes adjacent to an edge, divided by the total number of the nodes' connections  (excluding direct connections between the nodes \cite{Sporns2002}). A low MI identifies connections between very dissimilar network nodes which might represent important 'short cuts' between remote components of the network. 

Finally, edge frequency (EF), a measure similar to 'edge betweenness' \cite{Girvan2002,Holme2002b}, indicates how many times a particular edge appears in all pairs shortest paths of the network.  This measure focuses on connections that may have an impact on the characteristic path length by their presence in many individual shortest paths.  We used a modified version of Floyd's algorithm \cite{Cormen2001} to determine the set of all shortest paths and calculate the frequency of each edge in it. Multiple shortest paths between nodes $i$ and $j$ were present in the analyzed data sets. However, the standard algorithm only takes into account the first shortest path found. In order to account for edges in alternative shortest paths, the EF was calculated as the average of 50 node permutations in Floyd's algorithm. This led to an increased predictive value of this measure in all networks; however, the correlations already converged after 10 permutations. 

Another possible prediction measure, not used here, would be the range of an edge \cite{Watts1999,Sporns2002}, that is, the length of the shortest path between two adjacent nodes after the edge between them is removed. For dense networks, such as cortical connectivity, only range values of 2 and 3 occurred. Having only two classes of range values was not sufficient to distinguish vulnerable edges in detail. However, the range may be a useful predictor for sparse networks with higher ASP.

\section{Results}
\subsection{Network patterns underlying edge vulnerability}
The elimination of an edge from a network can have two possible effects on the ASP. First, the parts previously connected by this edge can still be reached by alternative pathways. If these are longer, the ASP will increase. Second, the eliminated edge may be a {\it cut-edge}, which means that its elimination will fragment the network into two disconnected components. The probability for fragmentation, naturally, is larger in sparse networks. Network separation causes severe damage, as interactions between the previously connected parts of the network are no longer possible. Therefore, this impact can be seen as more devastating than the first effect, which may only impair the efficiency of network interactions. Our ASP calculation disregarded paths between disconnected nodes, which would be assigned an infinite distance in graph theory. Therefore, the ASP in disconnected networks was actually shorter, because paths were measured within the smaller separate components. Cut-edges, which lead to network fragmentation, frequently occurred in the highway network (30\% of all edges) and the yeast protein interaction network (23\% of all edges). However, cut-edges did not occur for cortical networks of cat and macaque and only to a limited extent ($<$5\% of all edges) in the studied metabolic networks. 

In the present calculation both increase and decrease of ASP indicate an impairment of the network structure, we took the deviation from the ASP of the intact network as a measure for damage. We evaluated the correlation between the size of the prediction measures and the damage (Tab. \ref{corrtable} for all networks). While most of the local measures exhibited good correlation with ASP impact in real-world networks, the highest correlation was consistently reached by the EF measure. For the cortical networks, the measures of matching index and difference of degrees also show a high correlation.

Cortical connectivity differed from the other networks not only by the performance of different edge vulnerability predictors, but also in the density of connections and the amount of clustering. The cortical networks showed a higher density than the biochemical metabolic and protein-protein interaction networks. Whereas the highway network showed similar density to the biochemical networks, its clustering coefficient was much lower because of the high proportion of linear paths in the highway network. The random and scale-free benchmark networks --- designed to resemble size and edge density of the macaque cortical network --- are presented at the end of the table.

Fig. \ref{lincorr} shows the ASP after edge elimination plotted against edge frequency (EF). For cortical networks (\ref{lincorr} a,b) no network fragmentation occurred, and only an increase in ASP became apparent. For metabolic networks, e.g., the network of {\it S. cerevisiae} (\ref{lincorr} c), also a few cut-edges, lowering the ASP, were targeted. For the highway network containing linear chains of nodes (\ref{lincorr} d), many cut-edges were observed. The elimination of these links, therefore, resulted in two disconnected compartments, each of which had shorter path lengths. \\

\begin{figure}
\includegraphics{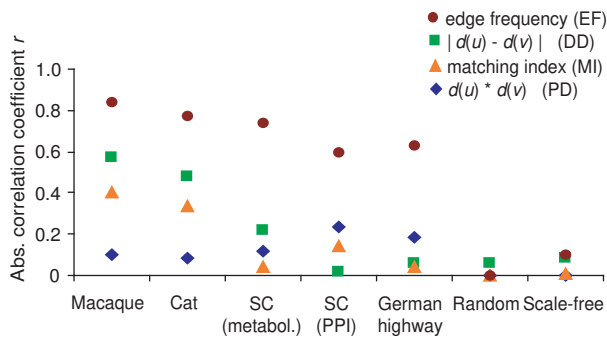}
\caption{Evaluation of the performance of four predictors for edge vulnerability. Note that the absolute correlation coefficient was used (MI would have had negative $r$). For all networks, except the random and scale-free benchmark networks, edge frequency had the highest correlation with edge vulnerability. In random networks and scale-free networks with only one cluster, however, the tested measures were unable to indicate impact of edge elimination. \label{corroverview}}
\end{figure}

\subsection{Comparison with benchmark networks}
\begin{figure}
\includegraphics{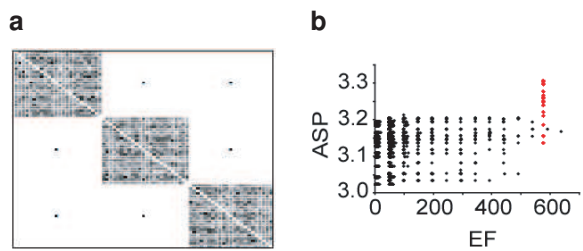} 
\caption{Connectivity for multi-clustered benchmark networks with comparable density to primate brain connectivity (cf. \ref{multibench}). The gray-level of a connection in the adjacency matrix indicates relative frequency of an edge in 20 generated networks. White entries stand for edges absent in all networks. \textbf{a} Connectivity of test networks with three clusters and six pre-defined inter-cluster connections. \textbf{b} Edge frequencies in the all-pairs shortest paths against ASP after elimination of edges. Red data points represent the values for the inter-cluster connections in all 20 test networks. Inter-cluster connections not only have the largest edge frequency, but also cause most damage after elimination.
\label{testmatrix}}
\end{figure}

We also calculated the four predictive indices for scale-free and random benchmark networks with 73 nodes and a similar number of edges as the macaque cortical network. Our comparisons focused on this network, because it showed the highest correlation between prediction measures and actual damage for all four measures. For the benchmark networks, however, all measures were poor predictors of network damage (Fig. \ref{corroverview}). This is surprising, because scale-free networks generated here by growth and preferential attachment appeared to differ in their structure from real scale-free networks. Analyzing data for one of these real networks, the Internet at the autonomous systems level, which was previously shown to be scale-free \cite{Barabasi1999}, we found that EF as a prediction measure was also performing well in this case ($r=0.62$, not shown). The difference between the real and simulated scale-free networks may result from the fact that scale-free networks generated by growth and preferential attachment did not possess multiple clusters. We therefore tested whether the lack of connections between clusters might be the reason for the low performance of EF in the scale-free benchmark networks. We generated further 20 test networks; each consisting of three randomly wired clusters and six fixed inter-cluster connections (Fig. \ref{testmatrix}a). The inter-cluster connections (light gray) occurred in many shortest paths (Fig. \ref{testmatrix}b) leading to an assignment of the highest EF value, as no alternative paths of the same length were available. Furthermore, their elimination resulted in the greatest network damage as shown by increased ASP. \\

\subsection{Network patterns in biological networks}
After establishing the high impact on ASP of edges with large EF, we investigated what made specific edges more vulnerable than others. We here discuss two patterns that occurred in many of the analyzed networks. First, linear chains of nodes that appeared in biological as well as the artificial (highway) networks. Second, clusters of highly interconnected regions of the network that occurred for all small-world networks, such as the analyzed cortical and biochemical systems. 

Naturally, other and more complex network patterns are possible. We already discussed elimination of edges between a hub and a node with fewer connections (cf. \ref{predictiondiscussion}). Also, the functional role of these patterns (e.g. feedback loops) was not examined here and would merit further study.

\subsubsection{Linear chains}
\begin{figure}
\includegraphics{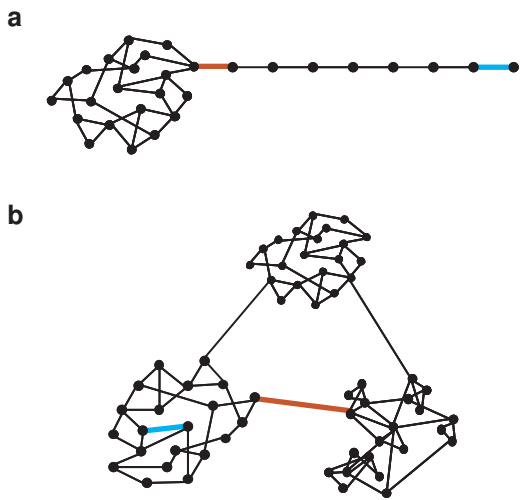} 
\caption{Network pattern and corresponding edge vulnerability. {\bf a} Elimination of edges of {\it linear chains of nodes} results in two disconnected components and a lower ASP. Edges eliminated at the proximal end of the chain (red) cause a larger change in ASP than at the terminal end (blue). {\bf b} For {\it clustered networks}, edges within the clusters (blue) can be replaced by several alternative pathways. Therefore, their elimination causes a smaller increase of ASP than that of edges between clusters (red).  \label{patterns}}
\end{figure}

Linear chains of nodes with a terminal end (Fig. \ref{patterns}a) became apparent in various biological as well as artificial networks. These patterns were detected by testing for each node if it was part of a chain, that is, if it possessed two edges. In this case the chain was followed in both directions, and considered terminal, if at least one end of the chain had a terminal node. Using this method for identification, each terminal chain consisted of at least two edges. Nodes in the terminal chain were excluded from the further searching process.  Terminal chains occurred frequently in the highway system, but also arose in metabolic networks in the form of redox chain reactions. For the highway system, the average terminal chain length was 6.4 edges, with a maximum of 22 edges. For the yeast protein interaction network, 13\% of the nodes were part of terminal chains, which were on average 2.25 edges long (maximum 6 edges). Eliminating edges at the terminal end of a chain would have a small impact, as only few nodes become disconnected to the rest of the network. On the other hand, severing the first edge that connects a chain to the rest of the network eliminates all paths leading to the chain nodes from the shortest paths matrix. The effect of eliminating edges in a chain can be seen clearly for the highway network (Fig. \ref{lincorr}d). Edges that connect chains to the rest of the network have a large EF, and their elimination greatly decreases ASP, in contrast to edges at the terminal end. Indeed, for networks that show many cut-edges, also many terminal chains occurred. For the highway system, 42\% of all nodes were part of terminal chains. Similar properties of edge vulnerability arise when the terminal of a chain end is formed by a small sub-network that is still smaller than the main network component at the start of the chain. \\

\subsubsection{Clustered architecture}
A clustered or modular architecture is a characteristic feature of many naturally occurring networks, such as cortical connectivity networks in the primate \cite{Young1992,Young1993,Hilgetag2000b} or the cat brain \cite{Scannell1995,Scannell1999,Hilgetag2000b} as well as metabolic networks \cite{Ravasz2002}. These systems are known to consist of several distinct, linked clusters with a higher frequency of connection within than between the clusters. Inter-cluster connections have also been considered important in the context of social contact networks, as 'weak ties' between individuals \cite{Granovetter1973} and separators of communities \cite{Girvan2002}. 
We, therefore, speculated that connections between clusters might be generally important for predicting vulnerability (Fig. \ref{patterns}b). Whereas many alternative pathways exist for edges within clusters, the alternative pathways for edges between clusters may be considerably longer.
Interestingly, previously suggested growth mechanisms for scale-free networks, such as preferential attachment \cite{Barabasi1999}, or strategies for generating hierarchical networks \cite{Barabasi2002} did not produce distributed, interlinked clusters. Consequently, the low predictive value of EF in the scale-free benchmark networks was attributable to the fact that scale-free networks grown by preferential attachment consisted of one central cluster, but did not possess a multi-cluster organization. This suggests that alternative developmental models may be required to reproduce the specific organization of biological networks, and we have recently presented an algorithm based on spatial growth that can generate such distributed cluster systems \cite{Kaiser2004b}.

\section{Discussion}
\subsection{Measures for identifying vulnerable edges\label{predictiondiscussion}}
We analyzed four measures for identifying vulnerable edges and predicting the impact of edge removal on global network integrity.  Among these measures, the index of EF appeared consistently as the best predictor for damage to edges. The high performance of this measure may be linked to characteristic features of biological networks, as detailed below (\ref{cluster}). For the macaque monkey, 7 out of the top 10 connections with highest edge frequency originated from, or projected to, the amygdala. In addition, these edges were most vulnerable as could be observed by the damage after edge elimination. Therefore, the amygdala appears to serve as a central link between many clusters of the network.

Following EF in terms of performance, the index for the difference of degrees also showed a high correlation in both the cortical and metabolic networks (Fig. \ref{corroverview}). This means that connections between highly and sparsely connected nodes are vulnerable, especially in cortical networks. When a node with few connections is connected to an already well-connected node (hub), it can access a large part of the network, by using routes involving the hub. After eliminating the connecting edge, the node would have to use longer alternative pathways to reach the same parts of the network. This effect is particularly strong, if the node was the only one in its local neighborhood that was connected to the hub. 

The matching index showed a large (negative) correlation with edge vulnerability in cortical networks and --- to a lower degree --- the protein-protein interaction network. Therefore, edges between dissimilar connections, that is, with low MI, were more vulnerable. This was due to the cluster structure of these networks (cf. \ref{cluster}). Nodes with similar connectivity belong to the same cluster, and therefore multiple alternative pathways are available. Dissimilar nodes are more likely to be part of different clusters with few alternative pathways.

From theoretical studies it has been proposed for scale-free networks that edges between hubs are most vulnerable \cite{Holme2002b}. However, in the networks analyzed here, both in the scale-free yeast protein interaction network as well as for cortical networks, edges which connected nodes possessing many connections (large product of degrees) were not particularly vulnerable. Although a low increase can be seen for the correlation coefficient, edges with maximum vulnerability occurred for a small product of degrees.

One of the general advantages of using prediction measures, instead of testing the damage for all edges of a network, is computational efficiency. For a network with $e$ edges and $n$ nodes, the order of time for the calculation of EF that can predict the effect of edge elimination for all edges is $\mbox{O(}n^3\mbox{)}$. This is lower than for testing the damage after edge elimination for all edges individually, calculating the ASP $e$ times resulting in a time complexity of $\mbox{O(}e\cdot n^3\mbox{)}$. The prediction measures presented here might therefore be particularly useful for large networks in which global testing is computationally impractical, or for networks with frequently changing connections demanding a regular update. Examples of these include acquaintance networks, Internet router tables, and traffic networks. Identifying vulnerable edges by EF generally appears to work well for various biological networks.

\subsection{Vulnerable edges in biological networks}
\label{cluster}
In the analyzed biological networks, inter-cluster projections may play an important role in linking functional units. For cortico-cortical networks, they connect and integrate different sensory modalities (e.g., visual, auditory) or functional sub-components \cite{Hilgetag2000b}. A lesion affecting these connections may result in dissociation disorders \cite{Geschwind1965}. 

For metabolic networks, reactions proceed more frequently within a reaction compartment (e.g., mitochondria and endoplasmatic reticulum) than between compartments. Therefore, in these systems as well, localized clusters arise, with many reactions within a compartment and few connections between compartments. Such an organization is also found in the investigation of protein-protein interactions and their spatial and functional clustering, in which fewer proteins from different groups interact \cite{Schwikowski2000}. Once again, interactions between proteins from different compartments correspond to inter-cluster connections, and might thus be among the most vulnerable edges of the network.

To a lower extent, 'inter-cluster' connections also appeared in the highway network, as the highway sub-systems of Western and Eastern Germany formed (spatially separate) dense regions connected by only four highways. That is, the elimination of four edges would once again split the German highway system into Eastern and Western components. Our analysis of artificial networks was restricted to networks without functional differentiation of edges or nodes. It remains to be seen if functional or social networks, for instance, interactions of people with different functions within a company, may show a higher similarity in cluster-architecture with biological networks. 

Clustered network architecture appears to result in edge-robustness in a similar way as scale-free architecture results in node-robustness. Random elimination of edges will most frequently select edges within a cluster. For paths routed through these edges, various alternative pathways exist, and the damage after edge elimination is small. Targeted attacks on inter-cluster connections, on the other hand, result in large network damage. We note that the cortical networks investigated here exhibit properties of both small-world \cite{Hilgetag2000b} and scale-free \cite{Kaiser2004e} networks and are therefore particularly robust to random failure of edges and nodes.

\subsection{Conclusions}
We examined how local network features and patterns relate to global network properties, such as robustness towards edge elimination. The correlation between different predictors and the actual damage after the elimination of a connection signified differences in the global network architecture. Specifically, various biological networks appear to be organized as distributed, linked network clusters. We have shown that inter-cluster connections represent the most vulnerable edges in these networks, and that their position can be predicted using the edge frequency measure.

\begin{acknowledgement}
M.K. acknowledges financial support from the German National Academic Foundation (Studienstiftung des deutschen Volkes).
\end{acknowledgement}

\bibliographystyle{apalike}

\begin{thebibliography}{10}

\bibitem[Albert et~al. 2000]{Barabasi2000a}
Albert R, Jeong H, Barab{\'a}si A-L (2000)
\newblock Error and attack tolerance of complex networks.
\newblock Nature 406:378--382

\bibitem[Barab{\'a}si and Albert 1999]{Barabasi1999}
Barab{\'a}si A-L, Albert R (1999)
\newblock Emergence of scaling in random networks.
\newblock Science 286:509--512

\bibitem[Barab{\'a}si et~al. 2001]{Barabasi2002}
Barab{\'a}si A-L, Ravasz E, Vicsek T (2001)
\newblock Deterministic scale-free networks.
\newblock Physica A 3--4:559--564

\bibitem[B\"uchel and Friston 1997]{Buchel1997}
B\"uchel C, Friston KJ (1997)
\newblock Modulation of connectivity in visual pathways by attention: cortical interactions evaluated with structural equation modelling and fMRI.
\newblock Cereb Cortex 7:768--778

\bibitem[Cormen et~al. 2001]{Cormen2001}
Cormen TH, Leiserson CE, Rivest RL, Stein C (2001)
\newblock Introduction to Algorithms.
\newblock MIT Press, Cambridge, Mass

\bibitem[Damier et~al. 1999]{Damier1999}
Damier P, Hirsch EC, Agid Y, Graybiel AM (1999)
\newblock The substantia nigra of the human brain. {II}. Patterns of loss of
  dopamine-containing neurons in Parkinson's disease.
\newblock Brain 122:1437--1448

\bibitem[Felleman and van Essen 1991]{Felleman1991}
Felleman DJ, van Essen DC (1991)
\newblock Distributed hierarchical processing in the primate cerebral cortex.
\newblock Cereb Cortex 1:1--47

\bibitem[Gavin et al. 2002]{Gavin2002}
Gavin et al. (2002)
\newblock Functional organization of the yeast proteome by systematic analysis of protein complexes.
\newblock Nature 415:141--147

\bibitem[Geschwind 1965]{Geschwind1965}
Geschwind N (1965)
\newblock Disconnection syndromes in animals and man: Part I.
\newblock Brain 88:229--237

\bibitem[Girvan and Newman 2002]{Girvan2002}
Girvan M, Newman MEJ (2002)
\newblock Community structure in social and biological networks.
\newblock Proc Natl Acad Sci 99:7821--7826

\bibitem[Granovetter 1973]{Granovetter1973}
Granovetter MS (1973)
\newblock The strength of weak ties.
\newblock Am J Sociol 78:1360--1380

\bibitem[Hilgetag et~al. 2000]{Hilgetag2000b}
Hilgetag CC, Burns GAPC, O'Neill MA, Scannell JW, Young MP (2000)
\newblock Anatomical connectivity defines the organization of clusters of cortical areas in the macaque monkey and the cat.
\newblock Phil Trans R Soc Lond Biol 355:91--110

\bibitem[Hilgetag et~al. 2002]{Hilgetag2002b}
Hilgetag CC, K{\"o}tter R, Stephan KE, Sporns O (2002)
\newblock Computational methods for the analysis of brain connectivity.
\newblock In: Computational Neuroanatomy, Humana Press, Totowa, pp 295--335

\bibitem[Holme et~al. 2002]{Holme2002b}
Holme P, Kim BJ, Yoon CN, Han SK (2002)
\newblock Attack vulnerability of complex networks.
\newblock Phys Rev E 65:056109

\bibitem[Ito et~al. 2001]{Ito2001}
Ito T, Chiba T, Ozawa R, Yoshida M, Hattori M, Sakaki Y (2001)
\newblock A comprehensive two-hybrid analysis to explore the yeast protein
  interactome.
\newblock Proc Natl Acad Sci, 98:4569--4574

\bibitem[Jeong et~al. 2001]{Jeong2001}
Jeong H, Mason SP, Barab{\'a}si A-L, Oltvai ZN (2001)
\newblock Lethality and centrality in protein networks.
\newblock Nature 411:41--42

\bibitem[Kaiser and Hilgetag 2004]{Kaiser2004b}
Kaiser M, Hilgetag CC (2004)
\newblock Spatial Growth of Real-World Networks.
\newblock Phys Rev E 69:036103

\bibitem[Kaiser et~al. 2004]{Kaiser2004e}
Kaiser M, Martin R, Andras P, Young MP (2004)
\newblock Structural robustness of cortical networks.
\newblock (submitted)

\bibitem[Keinan et~al. 2004]{Keinan2004}
Keinan A, Sandbank B, Hilgetag CC, Meilijson I, Ruppin E (2004)
\newblock Fair attribution of functional contribution in artificial and biological networks. 
\newblock Neural Comput (in press)


\bibitem[Kitano 2002]{Kitano2002}
Kitano H (2003)
\newblock Computational systems biology.
\newblock Nature 420:206--210

\bibitem[Ravasz et~al. 2002]{Ravasz2002}
Ravasz E, Somera AL, Mongru DA, Oltvai ZN, Barab{\'a}si A-L (2002)
\newblock Hierarchical organization of modularity in metabolic networks.
\newblock Science 297:1551--1555

\bibitem[Scannell et~al. 1995]{Scannell1995}
Scannell JW, Blakemore C, Young MP (1995)
\newblock Analysis of connectivity in the cat cerebral cortex.
\newblock J Neurosci 15:1463--1483

\bibitem[Scannell et~al. 1999]{Scannell1999}
Scannell JW, Burns GA, Hilgetag CC, O'Neil MA, Young MP (1999)
\newblock The connectional organization of the cortico-thalamic system of the cat.
\newblock Cereb Cortex 9:277--299

\bibitem[Schuster and Hilgetag 1994]{Schuster1994}
Schuster S, Hilgetag C (1994)
\newblock On elementary flux modes in biochemical reaction systems at steady
  state.
\newblock J Biol Systems 2:165--182

\bibitem[Schwikowski et~al. 2000]{Schwikowski2000}
Schwikowski B, Uetz P, Fields S (2000)
\newblock A network of protein-protein interactions in yeast.
\newblock Nature Biotech 18:1257--1261

\bibitem[Sporns 2002]{Sporns2002}
Sporns O (2002)
\newblock Graph theory methods for the analysis of neural connectivity
  patterns.
\newblock In: Neuroscience Databases, Kluwer Academic, Dordrecht, pp 169--183

\bibitem[Sporns et~al. 2000]{Sporns2000}
Sporns O, Tononi G, Edelman GM (2000)
\newblock Theoretical neuroanatomy: Relating anatomical and functional connectivity in graphs and cortical connection matrices.
\newblock Cereb Cortex 10:127--141

\bibitem[Stelling et~al. 2002]{Stelling2002}
Stelling J, Klamt S, Bettenbrock K, Schuster S, Gilles ED (2002)
\newblock Metabolic network structure determines key aspects of functionality
  and regulation.
\newblock Nature 420:190--193

\bibitem[Strogatz 2001]{Strogatz2001}
Strogatz SH (2001)
\newblock Exploring complex networks.
\newblock Nature 410:268--276

\bibitem[Wagner 2000]{Wagner2000}
Wagner A (2000)
\newblock Robustness against mutations in genetic networks of yeast.
\newblock Nature Genetics 24:355--361

\bibitem[Watts 1999]{Watts1999}
Watts DJ (1999)
\newblock Small Worlds.
\newblock Princeton University Press, Princeton

\bibitem[Watts and Strogatz 1998]{Watts1998}
Watts DJ, Strogatz SH (1998)
\newblock Collective dynamics of 'small-world' networks.
\newblock Nature 393:440--442

\bibitem[You et~al. 2003]{You2003}
You SW, Chen B, Liu H, Lang B, Xia J, Jiao X, Ju G (2003)
\newblock Spontaneous recovery of locomotion induced by remaining fibers after
  spinal cord transection in adult rats.
\newblock Restor Neurol Neurosci 21:39--45

\bibitem[Young 1992]{Young1992}
Young MP (1992)
\newblock Objective analysis of the topological organization of the primate
  cortical visual system.
\newblock Nature 358:152--155

\bibitem[Young 1993]{Young1993}
Young MP (1993)
\newblock The organization of neural systems in the primate cerebral cortex.
\newblock Phil Trans R Soc 252:13--18

\end{thebibliography}

\newcommand{\noopsort}[1]{} \newcommand{\printfirst}[2]{#1}
  \newcommand{\singleletter}[1]{#1} \newcommand{\switchargs}[2]{#2#1}

\end{document}